\documentstyle[prb,aps,multicol]{revtex}

\begin{document}

\preprint{Preprint}
\draft{}
%\twocolumn[\hsize\textwidth\columnwidth\hsize
%           \csname @twocolumnfalse\endcsname

\title{Observation of vortex lattice melting in large untwinned
YBa$_2$Cu$_3$O$_{7-x}$ single crystals}

\author{C.~M.~Aegerter\cite{Adr}, H.~Keller
}
\address{Physik-Institut der Universit\"at Z\"urich,
CH-8057 Z\"urich, Switzerland.}

\author{S.~H.~Lloyd, P.~G.~Kealey,
E.~M.~Forgan,
S.~T.~Johnson\cite{AAd}, and
T.~M.~Riseman
}
\address{School of Physics and Astronomy, University of Birmingham,
 Birmingham B15 2TT, UK.}

\author{R.~Cubitt
}
\address{Institut Laue Langevin, Grenoble France.}

\author{S.~L.~Lee, C.~Ager
}
\address{School of Physics and Astronomy, University of St.~Andrews,
St.~Andrews, Fife KY169SS, UK.}

\author{D.~M$^c$K.~Paul
}
\address{Department of Physics, University of Warwick, Coventry CV4 7AL, UK.}

\author{I.M. Savi\'c}
\address{Faculty of
Physics, University of Belgrade, 11001 Belgrade, Yugoslavia}

\author{M.~Yethiraj}
\address{Solid State Division, Oak Ridge National Laboratory, Tennessee USA.}

\author{S. Tajima and A. Rykov}
\address{Superconductivity Research Laboratory, ISTEC, Tokyo 135, Japan.}

\date{\today}
\maketitle
\widetext

\begin{abstract}
We present a study of the vortex lattice in untwinned YBa$_2$Cu$_3$O$_{7-x}$
crystals, using a combination of muon spin rotation and neutron small
angle scattering measurements. Both methods show a very sharp
melting temperature consistent with a first order transition. The
dependence of the melting temperature on the angle of the field with
respect to the crystallographic $c$-axis is studied. The results are
compared to thermal measurements.
\end{abstract}
\pacs{DRAFT VERSION: NOT FOR DISTRIBUTION}
%\pacs{PACS numbers: 74.25.Dw, 74.72.Bk, 74.60.Ge, 74.40.+k}

\begin{multicols}{2}
\narrowtext
%]
\section{Introduction}
Due to their short coherence length, long penetration depth and high
anisotropy, the high-T$_c$ superconductors (HTS), show very unusual
behaviour. Apart from a proposed vortex-glass phase \cite{fisher}, which
has probably been observed in the highly anisotropic
Bi$_2$Sr$_2$CaCu$_2$O$_8$ (BSCCO) at high fields \cite{lee93}, there is
also the
exciting phenomenon of vortex-lattice melting in these compounds. Unlike in
BSCCO, the vortex lattice melting transition in
YBa$_2$Cu$_3$O$_{7-x}$ (YBCO) takes place at sufficiently high fields
to be studied using thermal measurements such as
differential thermal analysis (DTA) \cite{schilling96} and specific heat
\cite{roulin96}. This is mainly due
to the much lower anisotropy of YBCO ($\gamma = \lambda_c / \lambda_{ab}
\simeq$ 5). Apart from this phenomenological difference, there are
however also more fundamental differences between YBCO and BSCCO in the
crystallographic symmetry, which is orthorhombic in the former, while it
is tetragonal in the latter. This leads to a fundamental difference in
the growth of the samples, in that untreated YBCO samples are usually
heavily twinned (having domains of interchanged $a$ and $b$ axes). The
boundaries between these domains (twin-planes) act as very strong pinning
sites to the vortices \cite{blatterbible}, which may change some of the
properties of the vortex lattice in the sample. It has for instance been
argued that such strong pinning may change the (intrinsically) first
order transition to a second order transition \cite{fisher}. This was
also supported by the fact that the transitions observed (for instance by
muon-spin rotation ($\mu$SR) \cite{lee93} and neutron small-angle
scattering (SANS) \cite{cubittnature}) in BSCCO were
generally very sharp, not inconsistent with a first order transition,
whereas twinned YBCO samples gave a more gradual transition. With the
advent of sufficiently sized untwinned and high purity twinned samples,
it became possible to study these effects in more detail using
DTA\cite{schilling96} and
specific heat\cite{junod} to determine the true thermodynamic properties of
the
phase transition. In untwinned samples, DTA measurements have clearly
shown the existence of a first order transition\cite{schilling96}, even
though the sample
wasn't pure but had some Au impurities due to the growth process. In
extremely pure samples, specific heat measurements have also shown a
first order transition even though the sample was heavily
twinned\cite{junod}.
However, the twinned case is in general much more complicated than the
untwinned one in that the order of the transition may also depend on the
applied field\cite{junod}. The order of the transition is thus probably
influenced by
both kinds of defects (impurities and twin planes), where an excess of
pinning sites of any kind drives the transition to second order.

Of further interest is the dependence of the melting transition on the
angle of the field with respect to the $c$-axis \cite{schilling97}. There
are detailed predictions on the field dependence of the
melting temperature (melting line) on the angle from anisotropic London
theory \cite{blatterani}. The nature of the transition, especially the
change in entropy at melting has however only recently been investigated
\cite{schilling97,dodgson}.

We have now performed $\mu$SR and SANS measurements of the vortex lattice
in a large untwinned crystal of YBCO with the field at different angles
to the crystallographic $c$-axis. As is the case in the DTA and
specific heat measurements, we probe the {\em equilibrium state} of the
vortex lattice. From SANS, we obtain a microscopic measurement of the
global (long range) arrangement of the flux lines. Appropriately scaled,
the SANS intensity presents a phenomenological order parameter of the
flux lattice (see below). This allows investigations similar to those of
specific heat or DTA, clarifying the order of the transitions. It should
however be noted that the samples used in SANS are very much bigger than
those used for thermal measurements. Thus a possibly present first order
transition may already be smeared by slight inhomogeneities over the big
sample volume. In contrast, with $\mu$SR we observe the local magnetic field
structure averaged over the sample. Therefore smearing effects may still
be present, however a sharp change of the local equilibrium structure in
itself may not be regarded as a sign of a first order transition. Due to the
different time scales available to neutrons and muons one would expect to
observe a difference in the melting line determined by the two methods if
the transition is of a more glassy nature. For a (first order) phase
transition, one would expect both methods of measurement to agree.

As will be shown below, our results differ quite strongly from those on a
heavily twinned sample \cite{aegerter98,roulin}. There the melting transition
observed with SANS is of second order, with the intensity decreasing to
zero continuously at a temperature T$_m <$ T$_c$. In contrast our results
show a sharp decrease of the SANS intensity with an infinite slope at
T$_m$. This is consistent with a first order transition considering the
large size of the sample (see below).

\section{Experimental Details}
The sample consisted of a large (1.125 g) YBCO single crystal. It was grown
untwinned by application af uniaxial stress during the cooling from the
melt \cite{rykov}.
For the oxygenation process, the sample was treated in flowing O$_2$ at
490$^\circ$C. Magnetisation measurements using a vibrating sample
magnetometer (VSM) indicate a T$_c$ of 93 K. The superconducting
transition is estimated from these measurements to be $\sim$0.5 K.
This also gives a limit
for the sharpness of the melting transition (see below).

The $\mu$SR experiments were carried out on beamline $\pi$M3 at the Paul
Scherrer Institute (PSI), Villigen Switzerland, with a maximum field of
0.6 T and a cryostat capable of cooling to 2 K. In the experiments, a
magnetic field was applied perpendicular to the incoming muon spin
polarisation, while at a specified angle to the crystallographic
$c$-direction of the sample. In this case, the spin of a stopped muon
precesses with the frequency $\omega = \gamma_\mu$B$_{loc}$, where
$\gamma_\mu = 2\pi$ 135 MHz/T is the gyromagnetic ratio of the muon in
the transverse local field B$_{loc}$. The angular anisotropy of the
muon decay, leading to a preferred emission of the decay-positron in the
direction of the muon spin due to parity violation, allows the time
evolution of the muon polarisation to be measured. In the mixed state of
a type-II superconductor, the magnetic field distribution is given by the
structural properties of the vortex lattice (i.e. the penetration depth
$\lambda$, the coherence length $\xi$ and the lattice morphology). The
vortex lattice spacing is much bigger than the atomic spacing of the
underlying crystal
structure and is incommensurate to it. Therefore the muons sample the
field distribution randomly, even though they are stopped at specific
crystallographic sites. This allows the field probability distribution
p(B) to be measured via the Fourier transform of the oscillating time
signal. As the different local fields cause the muon spin to precess at
different frequencies the time signal consists of a sum of oscillations
at different frequencies weighted by their relative occurrence. In order
to obtain the Fourier transform from the observed time data, we use a
Maximum Entropy algorithm \cite{rainford}. In this procedure aberrations
of the spectra
due to finite time windows or large statistical errors at long times are
avoided, as the data are not subjected to a Fourier transform themselves.
Starting from an initially flat field distribution p(B), the most uniform
distribution compatible with the given time spectrum is obtained by
maximising its constrained entropy
\begin{equation}
L = S - \lambda \chi^2.
\end{equation}
Here $S = \sum p(B) ln(p)$ is the entropy of the distribution according to
Shannon and $\chi^2 = \sum (y_i - d_i)^2/\sigma^2$ is the usual measure
giving the difference of the Fourier transform of p(B) ($y_i$) to the
data in the time domain ($d_i$), where $\sigma$ is the error in the data.

The field distribution corresponding to a vortex lattice has very
characteristic features arising from its two-dimensional periodicity. At
extremal points in the spatial distribution the field probability
distribution exhibits van Hove singularities. These singularities ideally are
finite jumps at the minimum and maximum fields, and furthermore a
logarithmic divergence at saddle points. Due to the very high fields in
the vortex cores, the field distribution is highly asymmetric with a long
`tail' at fields higher than the average. In the entangled vortex liquid state
however, the core fields are smeared out over a large distance. Therefore
the high field tail is truncated leading to a symmetric lineshape. We
quantify the asymmetry of the lineshape by the dimensionless parameter
\begin{equation}
\alpha = \frac{\langle \Delta B^3\rangle^{1/3}}{\langle \Delta B^2
\rangle^{1/2}}.
\end{equation}
A symmetric distribution corresponds to a value of $\alpha = 0$, whereas
a distribution with a high field tail has a positive $\alpha$.

The SANS experiments were carried out on instruments D11, D22 and
D17 at the Institut Laue-Langevin (ILL), Grenoble France. The
neutrons typically had a wavelength of $\lambda_n$ = 1 - 1.5 nm
and the beam was collimated over a distance of 2.75 m in the
experiments carried out at D17, whereas the collimation length was
10 m in those at D11 and D22. The distance between the sample and
the detector was then varied for optimal resolution. The incoming
neutron beam was aligned with the applied field to an accuracy of
0.1$^\circ$ by observing the diffraction pattern from a Nb single
crystal. The sample was mounted in a cryomagnet capable of a field
of 5 T and a temperature of 1.5 K.

The neutrons are scattered from the spatial variations of the magnetic
field in the sample, due to the magnetic moment of the neutron ($\mu_n =
-1.913 \mu_N$), with the interaction energy $\vec{\mu}_n \cdot \vec{B}$. In
the case of a type-II superconductor in the mixed state, the field
variation is periodic giving rise to Bragg diffraction off the
scattering planes of the vortex lattice \cite{forgannature,nb}, where the
angle of refraction is given by
\begin{equation}
\lambda_n = 2 d_{hk} sin(\theta).
\label{bragg}
\end{equation}
Here $d_{hk}$ is the lattice spacing, which depends on the applied field
as $d_{hk} = (\sqrt{3}/2 \Phi_0 /B)^{1/2}$ for a hexagonal lattice
($\Phi_0$ = h/2e is the magnetic
flux quantum). Thus the scattering is at very small angle, as typical
vortex separations of $\sim$50 nm (corresponding to a field of $\sim$1
T) result in a Bragg angle (from Eq.~\ref{bragg}) of 2$\theta \simeq$
2$^\circ$. The intensity of neutrons scattered into a Bragg-peak,
integrated over angle as the sample is rocked through the Bragg
condition, is then mainly given by the square of the form factor of a
flux line, which in the London approximation is given by
\begin{equation}
F(q) = \frac{B}{1 + \lambda^2 q^2},
\end{equation}
where $\lambda$ is the penetration depth. The London approximation is
valid in the case of an extreme type-II superconductor with $\lambda \gg
\xi$, a condition well satisfied by the HTS. Therefore, for fields much
greater than B$_{c_1}$, corresponding to
$\lambda q \gg$ 1, the scattered intensity depends on the penetration
depth as $I \propto \lambda^{-4}$. For a classical type-II
superconductor, the temperature dependence of the scattered intensity
therefore is a direct measure of that of the penetration depth. In the
HTS however, where thermal fluctuations may have a dominant contribution
this is not necessarily the case. In order to discern effects arising from the
temperature dependence of $\lambda$ and those of the thermal fluctuations
finally leading
to the melting of the lattice we divide the observed intensities by the
temperature dependence of $\lambda^{-4}$. This was obtained from the SANS
intensity at a low field (0.2 T), where the effects of thermal
fluctuations are negligible (see below). This quantity then takes the
role of an effective flux line lattice (FLL) order parameter, which we use to
observe the melting transition and its order.

Because the scattered intensity depends on the penetration depth as
$I \propto \lambda^{-4}$ the intensity in the HTS is rather low, due
to their long penetration depths
($\lambda$ = 150 - 200 nm). Furthermore, substantial background scattering
from the cryostat and extended defects in the sample make it necessary to
subtract a diffraction pattern taken above T$_c$ to obtain a clear signal.
Due to thermal contractions of the sample stick and cryostat,
this background is shifted by fractions of a detector pixel in order to
obtain a good subtraction. Thanks to the high neutron intensity at the
instruments D22 and D11, a counting time of 30 minutes for each foreground
and background was enough to obtain a reasonable signal.

\section{Results and Discussion}
\begin{figure}
\input{epsf}
\epsfysize 6.5cm \centerline{\epsfbox{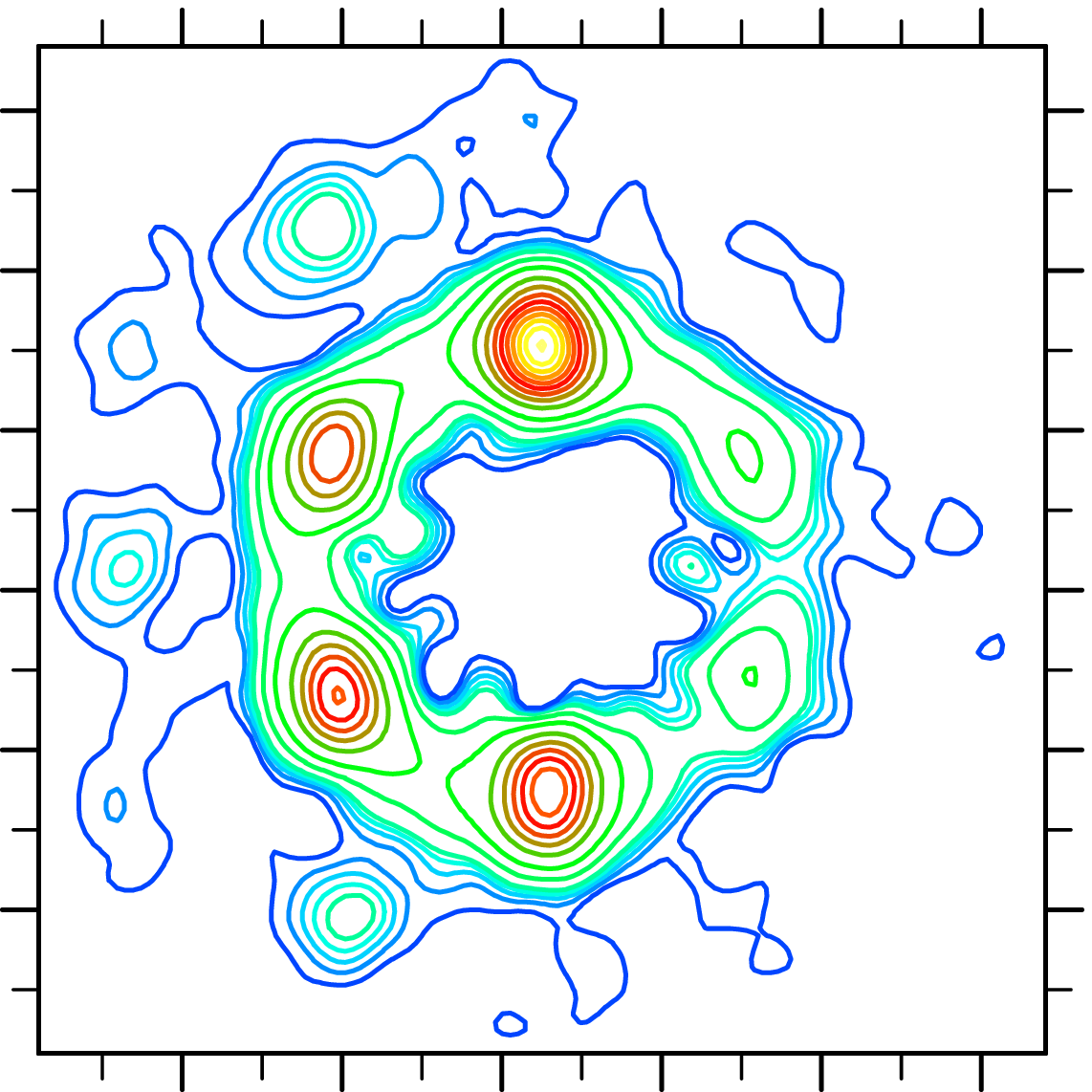}} \caption[~]{A
typical SANS diffraction pattern from the vortex lattice in YBCO.
A field of 0.2 T was applied at 33$^\circ$ to the crystallographic
$c$-axis and the sample was cooled through T$_c$ in the field to a
temperature of 5 K where the image was taken. In order to remove a
background of small angle scattering, a diffraction pattern taken
above T$_c$ was subtracted. Note that the diffraction pattern is
an undistorted hexagon due to the fact that the field is at
33$^\circ$, undoing the in-plane anisotropy $\gamma_{ab}$.
Furthermore, second order diffraction spots are visible in the
figure, indicating the high degree of order in the flux lattice. }
\label{pattern}
\end{figure}

A typical diffraction pattern can be seen in Fig.~\ref{pattern} for an
applied field of 0.2 T. There
the field is applied at an angle of 33$^\circ$ to the $c$-axis. This is
why the observed pattern is isotropic. The distortion due to the
$ab$-anisotropy ($\gamma_{ab}$ = 1.16 \cite{stu,cam}) is compensated by the
distortion due to the $ac$-anisotropy at an angle of 33$^\circ$.
Furthermore, we note that second order diffraction spots are visible in
the figure, indicating the high quality of the flux lattice. The
intensities measured from such a pattern are obtained from fitting a
gaussian to each individual Bragg-spot and taking the sum of all spots
in order to reduce the statistical error. All of these measurements of
scattered intensity rely on the fact that the width of the rocking curve
does not change as a function of temperature. This is because the
calculation relating the scattered intensity to the penetration depth is
done for the total integrated intensity\cite{mook}, which is given by the
width of
the rocking curve multiplied by the intensity when fulfilling the Bragg
condition. The rocking curve width was checked to be constant for selected
temperatures over a wide range. We therefore only consider the scattered
intensities at the Bragg-angle.

\begin{figure}
\input{epsf}
\epsfysize 6.5cm \centerline{\epsfbox{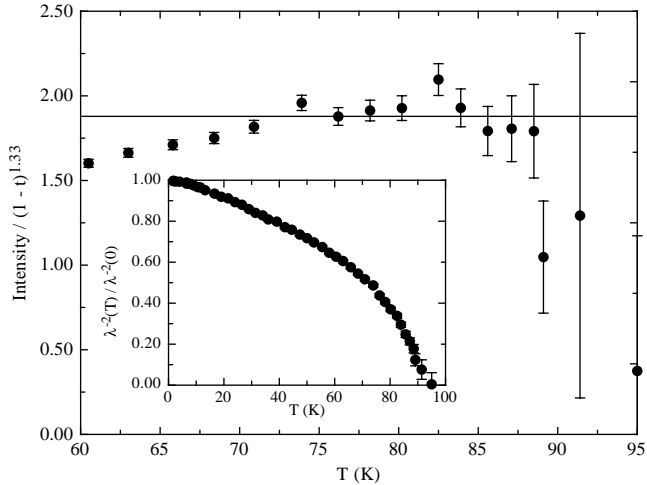}} \caption[~]{The
temperature dependence at a low field of 0.2 T gives a measure of
the penetration depth. For clarity, the scattered intensity has
been divided by (1 - T/T$_c$)$^{1.33}$, which would be the
expected temperature dependence from the 3D XY model. This is
thought to be applicable in the vicinity of T$_c$. Our data
however show that even at temperatures as low as 75 K the
approximation is still valid. The insert shows the full
temperature dependence of $\lambda^{-2}$, indicating a linear
variation at low temperatures. The saturation at very low
temperatures may be due to non-local effects expected in a d-wave
superconductor\cite{kosztin} (see the text). } \label{tdep}
\end{figure}

In order to determine an effective FLL order parameter we have to divide
the intensity by its temperature dependence intrinsically
given by the penetration depth. Therefore we have measured
the scattered intensity at a low field (0.2 T) where the influence of
thermal fluctuations should be small. The result can be seen in
Fig.~\ref{tdep}, where the intensity at 0.2 T is given divided by the
expectation of the 3D XY model. As the figure shows, the approximation of
the 3D XY model, with $\lambda \propto$ (1 - T/T$_c$)$^{-0.33}$ is valid
for temperatures well below T$_c$ in all of the region of interest. In
the determination of the order parameter, the intensity was thus divided
by (1 - T/T$_c$)$^{1.33}$ for simplicity. Furthermore the insert of
Fig.~\ref{tdep} shows the temperature dependence over the full range.
There we have plotted the square root of the scattered intensity to
present the familiar $\lambda^{-2}$ dependence \cite{muwave}. It can be seen
that at lower temperatures the dependence is linear down to temperatures of
$\sim$4 K, below which it saturates. For a d-wave superconductor, as
the HTS are supposed to be, a linear temperature dependence would be
expected down to the lowest temperatures from simple London theory. However
when non-local interactions are taken into
account \cite{kosztin}, a saturation at very low temperatures is
predicted, not inconsistent with our data. We will come back to the
temperature dependence of $\lambda$ below in the discussion of the melting
line. However, we note that such a big range of applicability (down to
$\sim$20 K below T$_c$) for the 3D
XY model is surprising, as this is thought only to be valid in the region
of critical fluctuations in close vicinity of T$_c$.

\begin{figure}
\input{epsf}
\epsfysize 6.5cm \centerline{\epsfbox{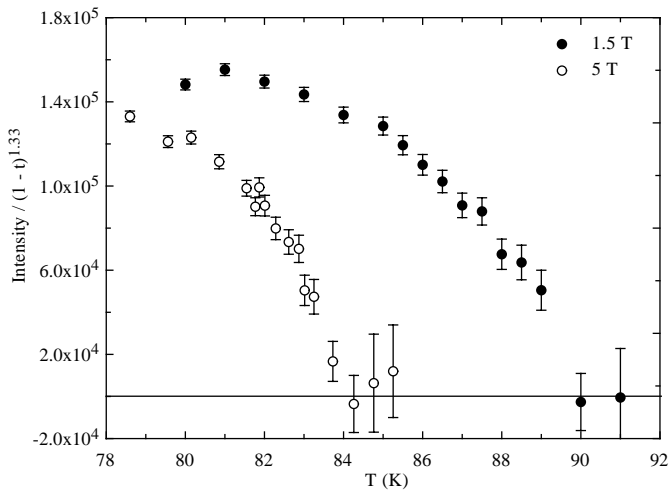}}
\caption[~]{The temperature dependence of the FLL order parameter
at two different fields of 1.5 and 5 T. In both cases, the order
parameter decreases to zero very rapidly at the melting
transition. This is consistent with the transition being of first
order (see the text). Furthermore the fact that the scattered
intensity is zero above the melting temperature indicates the
presence of an entangled vortex liquid. } \label{SANSmelt}
\end{figure}

In Fig.~\ref{SANSmelt}, we present the temperature dependence of the FLL
order parameter in the vicinity of T$_c$ for two different fields. At
both fields, the order parameter is constant at low temperature and
decreases gradually towards the melting temperature. When the order
parameter finally drops to zero, the change is very sharp occurring over a
temperature range of less than 1 K. The gradual decrease of the order
parameter can be modeled by including a Debye-Waller factor into the
theoretical description of the scattered intensity. The final `jump' however,
is a sign of the transition being
of first order. The width of the transition of 1 K is not  much bigger
than that observed at T$_c$, which may be due to sample inhomogeneities.
Such inhomogeneities leading to a distribution of T$_c$ may also lead to
a distribution of T$_m$ which would probably be of the same order of
magnitude. Furthermore, the distribution of fields inside the sample is
inhomogenious \cite{indenbom}, as we will also see in the discussion of
the $\mu$SR results below, which will also lead to a further smearing of the
melting temperatures. However, as this macroscopic field distribution is very
narrow compared to the applied fields, this shouldn't be a big factor.
It is therefore not unreasonable to expect the
melting transition in such a big sample to be distributed over an
interval of $\sim$1 K. Note also that the order parameter (the scattered
intensity) decreases to zero above the melting transition, in agreement
with the expectation for an entangled vortex liquid
\cite{santacruz,nordborg} made up of two-dimensional `pancake' vortices.

\begin{figure}
\input{epsf}
\epsfysize 6.5cm \centerline{\epsfbox{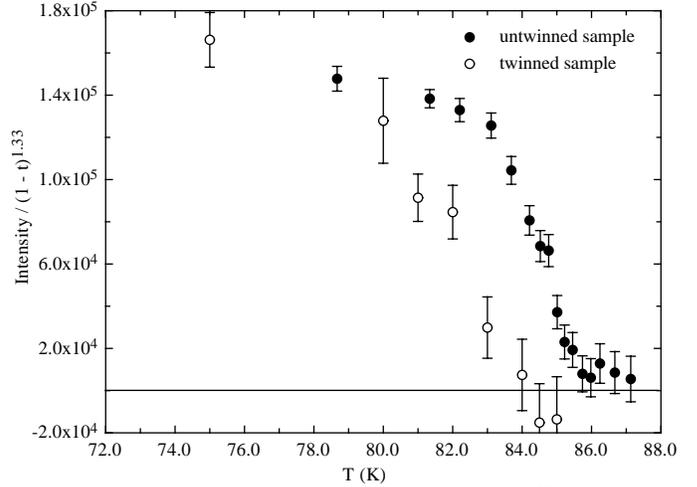}} \caption[~]{The
FLL order parameter in a twinned \cite{aegerter98} and in our
untwinned sample in an applied field of 4 T. In the untwinned
sample, the melting transition is step-like with a width of
$\stackrel{<}{_\sim}$1 K. In the twinned sample in contrast, the
order parameter decreases continuously over more than 5 K below
the melting temperature. Close to T$_m$, the decrease in the
twinned sample is linear, as would be expected for an order
parameter in mean field theory. } \label{comp}
\end{figure}

The situation for the untwinned sample has to be compared to that in a
heavily twinned sample \cite{aegerter98}. This is done in
Fig.~\ref{comp}, where we show the FLL order parameter for both a twinned
and our untwinned sample in a field of 4 T. As can be seen in the
figure, the transition is much sharper in the untwinned one. The
order parameter decreases more or less linearly over a range of 5 K in
the twinned sample, as would be expected from mean field theory. In the
untwinned sample in contrast, the transition is at most 1 K wide. Takeing
also into account the temperature dependence at lower temperature, where
the untwinned sample varies much slower than the twinned one, the
behaviour of the untwinned sample is consistent with a first order
transition. The width of the observed transition is furthermore not very
much bigger than those determined from magnetisation\cite{welp} and
thermal measurements\cite{schilling97},
which are done on much smaller samples where inhomogeneities and possible
residual twin-planes must play a much smaller role.

\begin{figure}
\input{epsf}
\epsfysize 6.5cm \centerline{\epsfbox{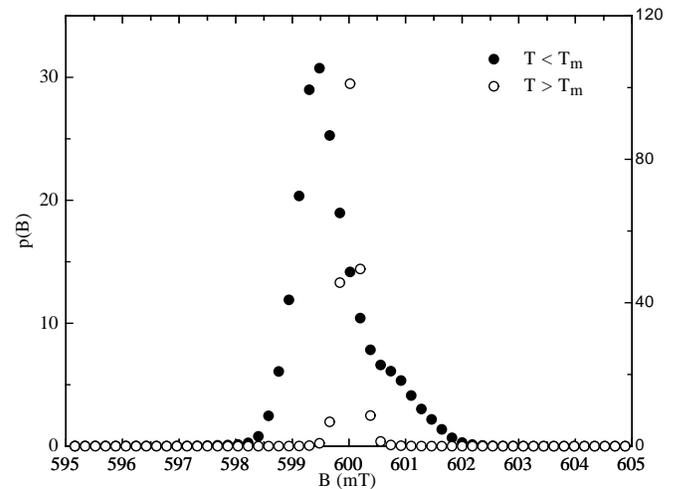}}
\caption[~]{The $\mu$SR lineshapes above and below the melting
temperature in an applied field of 0.6 T. At low temperatures,
the field distribution is asymmetric towards high fields, as is
expected for a lattice of straight vortex lines due to the high
local fields in the vortex cores. Above T$_m$ however, the
lineshape is considerably narrowed due to thermal fluctuations.
The expected symmetric shape is changed to a weighting towards
low fields because of a macroscopic distribution of fields over
the whole of the sample (see the text). } \label{lineshape}
\end{figure}

We now turn to the discussion of the $\mu$SR results. Typical field
probability distributions for temperatures above and below the melting
transition can be seen in Fig.~\ref{lineshape} for a field of 0.6 T. The
situation is similar
to that of BSCCO \cite{lee93,lee97}, with a lineshape highly asymmetric
towards high fields below the melting temperature. As was discussed
above, the lineshape for an ideal lattice made up of straight vortex
lines is expected to show such a high field tail. Above the melting
temperature however, the field distribution is skewed the opposite way.
This has been argued to arise from a macroscopic distribution of the
local fields due to the sharp edges of the sample, with fields decreasing
at the edges \cite{ted,indenbom}. This
kind of macroscopic distribution is also present below the melting
temperature,
however the large width of the field distribution due to the well formed
lattice overcomes this effect, such that an almost ideal distribution is
observed. Above the melting temperature in contrast, the field
distribution is smeared by the fast thermal fluctuations of the pancake
vortices.
This leads to a strongly decreases the width of the microscopic field
distribution (`motional narrowing')
and hence the observed distribution follows the macroscopic one.

\begin{figure}
\input{epsf}
\epsfysize 6.5cm \centerline{\epsfbox{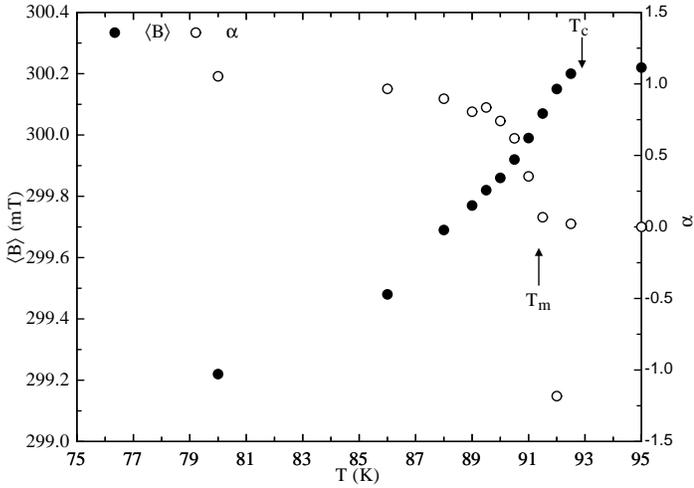}}
\caption[~]{The temperature dependence of the mean internal field
$\langle B \rangle$ and the lineshape asymmetry parameter
$\alpha$ at a field of 0.3 T. Even at such a low field, there is
a sharp melting transition as evidenced by the change in sign of
$\alpha$. This takes place at a temperature below T$_c$ as
determined from the onset of diamagnetic behaviour in $\langle B
\rangle$. This change in $\alpha$ corresponds to that in
lineshape displayed in Fig.~\ref{lineshape} consisting in the
truncation of the high field tail due to a smearing of the core
fields because of the thermal fluctuations of the pancake
vortices. } \label{musrmelt}
\end{figure}

In Fig.~\ref{musrmelt}, we present the temperature dependence of the
$\mu$SR lineshape asymmetry parameter $\alpha$, together with that of the
mean internal field. The mean field allows a direct measurement of T$_c$
via the onset of diamagnetic behaviour. As can be seen in the figure even
at the very low field of 0.3 T, we observe a sharp melting transition in
the temperature dependence of $\alpha$ somewhat below T$_c$. The change
in lineshape displayed in Fig.~\ref{lineshape} manifests itself by a
change in sign of $\alpha$. The transition is still very sharp, although
thermal measurements are not sensitive enough to resolve a first order
transition at such low fields. There have however recently been
magnetisation measurements by means of torque magnetometry, resolving a step
in magnetisation and hence a first order transition at similarly low
fields \cite{michel}.

\begin{figure}
\input{epsf}
\epsfysize 6.5cm \centerline{\epsfbox{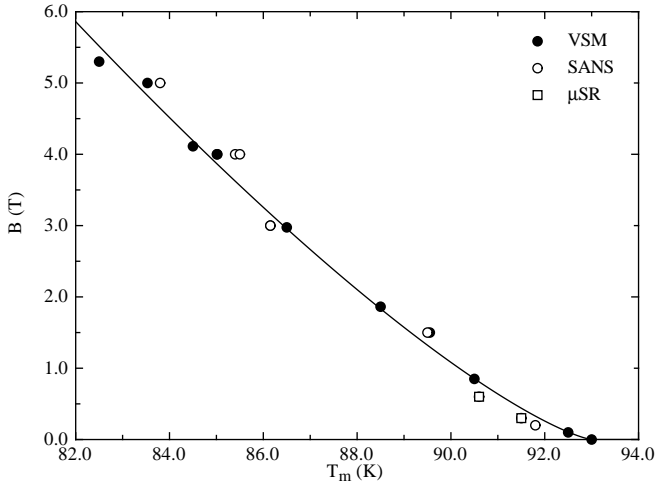}}
\caption[~]{The melting line as determined with the field
parallel to the $c$-direction. Filled circles represent the
irreversibility line as determined from the onset of irreversible
magnetisation using a VSM. Open circles are determined by SANS
from the jump in the FLL order parameter. The squares correspond
to $\mu$SR measurements, where the change in sign of $\alpha$ was
taken as an indication of the melting transition. The solid line
is  a fit to all of the data of a melting line of the form $B_m =
B_0$ (1 - T/T$_c$)$^n$ with $B_0$ = 100(10) T and n = 1.30(5).
This is in agreement with the expectations for a melting line
given the temperature dependence of $\lambda$ from
Fig.~\ref{tdep} (see the text). } \label{meltline}
\end{figure}

The field dependence of the melting temperatures thus determined (the
melting line) can be seen in Fig.~\ref{meltline}. In that figure we have
also included the line of onset of irreversible magnetisation
(irreversibility line, IL) as determined with a VSM. The microscopic
melting line and the IL are in good agreement with each other. They follow
a power law dependence of the form
\begin{equation}
B_m (T) = B_0 (1 - T / T_c)^n.
\end{equation}
A least squares fit to both sets of data simultaneously yields the values
of $B_0$ = 100(10) T and n = 1.30(5). This is in excellent agreement with
many other studies \cite{schilling96,roulin,aegerter98} on both twinned
and untwinned samples. It is thus only the order of the transition, which
is affected by the presence of disorder (e.g. in the form of twin-planes)
and not the meltingline itself. The shape of the melting line has been the
ground of some debate, as a dependence $B_{VG}\propto$ (1 -
T/T$_c$)$^{1.33}$ is expected for the vortex-glass transition proposed by
Fisher {\em et al.} \cite{fisher}. This would indicate that the observed
melting line is more likely to be a glass transition, which would however
be in contradiction with the observation of a first order transition. The
temperature dependence of a melting line is usually calculated using the
Lindemann criterion that melting occurs when the mean square
displacements of flux lines
make up a certain fraction of the intervortex separation $\langle u^2
\rangle = c_L^2 a_0^2$. The Lindemann number $c_L$ is supposed to be a
universal constant in all kinds of melting processes and theoretical
estimates indicate it to be 0.1 - 0.2. Using this criterion and modeling
the thermal fluctuations in the vortex lattice one can then derive an
expression for the melting line \cite{houghton}
\begin{equation}
B_m = \frac{\Phi_0^5 c_L^4}{(1.5 \pi \mu_0 \gamma k_B T \lambda^2(T))^2}.
\end{equation}
Due to the mean field assumption of the temperature dependence of the
penetration depth $\lambda^{-2}$ (T) $\propto$ (1 - T/T$_c$)$^{1/2}$ this
expression is usually taken to result in a value of n = 2. As we have
seen however in Fig.~\ref{tdep}, $\lambda$ follows a powerlaw,
such that the value of n = 1.33 is recovered over the range of
temperatures where the meltingline can be determined. Furthermore, this
expression allows a determination of $B_0$, where $c_L$ is the only adjustable
parameter. From the Intensity in the Bragg peaks, we have an independent
determination of $\lambda$ = 138 nm \cite{stu} and the anisotropy of our
sample has been determined from the angular dependence of the $\mu$SR
linewidth as well as from the distortion of the SANS pattern at high
angles to the $c$-direction. Therefore our value of $B_0$ = 100(10) T
corresponds to a Lindemann number of $c_L$ = 0.173(7). This is in
excellent agreement with that found from the melting line in BSCCO using
$\mu$SR \cite{lee97}. We would like to note at this point that the
Lindemann number determined in BSCCO relied on a different expression for
the thermal fluctuations taking into account the electromagnetic
coupling between pancake vortices. This indicates that the Lindemann
criterion may indeed be universally applicable.

\begin{figure}
\input{epsf}
\epsfysize 6.5cm \centerline{\epsfbox{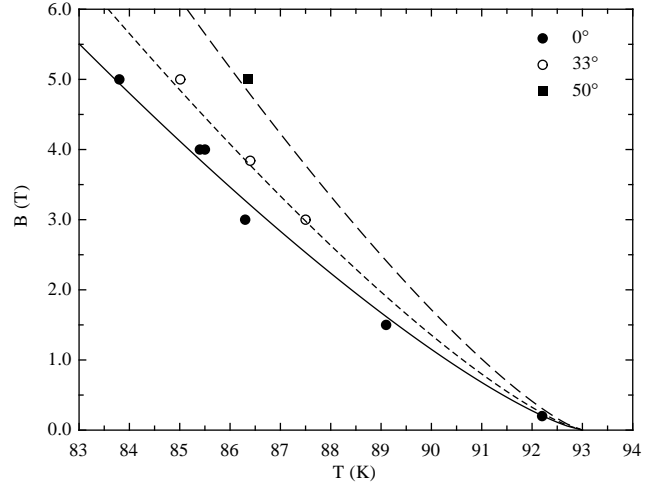}}
\caption[~]{The melting line as determined by SANS for different
angles with respect to the $c$-axis. The lines correspond to the
expectations from the anisotropic scaling theory of Blatter {\em
et al.} \cite{blatterani} given the melting line parallel to $c$
(see Fig.~\ref{meltline}) and an anisotropy of $\gamma$ = 4 (see
Ref. \cite{cam}). A determination of $\gamma$ from the melting
lines is not accurate enough, as the studied angles only extend
to 50$^\circ$. } \label{anglemelt}
\end{figure}

\begin{figure}
\input{epsf}
\epsfysize 6.5cm \centerline{\epsfbox{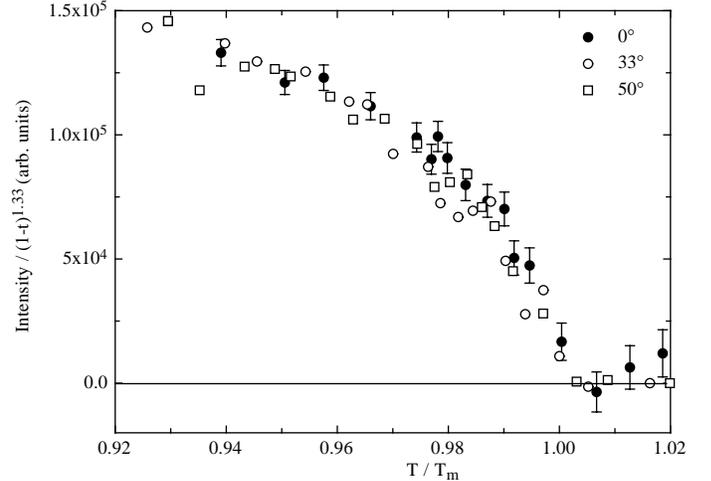}}
\caption[~]{The FLL order parameter in a field of 5 T at three
different angles to the $c$-direction. The temperatures are
scaled to the respective melting temperatures to give a unified
picture. Within the errors, the transitions are the same, which
is consistent with the finding of a constant entropy change as a
function of angle \cite{schilling97}. } \label{angledep}
\end{figure}

Finally, we discuss the dependence of the melting transition on the angle
of the applied field to the $c$-direction. From the anisotropic scaling
theory of Blatter {\em et al.} \cite{blatterani}, a dependence of the
melting line on angle is
\begin{equation}
B_m (\vartheta) = B_m (0) (cos^2(\vartheta) +
\gamma^{-2}sin^2(\vartheta))^{-1/2}.
\end{equation}
In the region of angles we investigated ($\vartheta <$ 50$^\circ$), the
influence of $\gamma$ on $B_m$ is very small such that it cannot be
accurately determined this way. However using the value obtained from
$\mu$SR and the distortion of the pattern \cite{cam}, we may calculate
the  expected melting lines at an angle from that determined parallel to
$c$. This has been done in Fig.~\ref{anglemelt}, where we show the
melting temperatures determined by SANS at the three different
angles of 0$^\circ$, 33$^\circ$ and 55$^\circ$. The lines correspond to
the fit to the melting line discussed above, where the dependence on the
angle has been taken into account as well. As can be seen in the figure,
the data are in good agreement with the expectations of the anisotropic
scaling theory. This is similar to the results of DTA measurements of the
melting transition as a function of angle \cite{schilling97}. In
addition, that work also provided a determination of the entropy change at the
transition as a function of angle. Due to a cancellation of the angular
dependencies in the melting line and the magnetisation jump, the
entropy-change turns out to be independent of the angle. This is not
inconsistent with the SANS result displayed in Fig.~\ref{angledep}, where
the temperature dependence of the FLL order parameter is shown at the
melting transition for different angles. The temperatures are normalised
to T$_m$ for direct comparison and the applied field was 5 T. The
decrease of the order parameter can be seen to be independent of the
angle as it should be for a constant entropy change at the transition.

\section{Conclusions}
In conclusion, we have presented SANS and $\mu$SR measurements of the
melting of the vortex lattice in a sample of untwinned YBCO. The observed
behaviour is consistent with DTA measurements of the latent heat showing
the existence of a first order transition \cite{schilling96,schilling97}.
This is inferred from a
phenomenological FLL order parameter in SANS, which decreases to zero
sharply at the melting transition. These findings are compared to a
previous SANS study of melting in twinned samples \cite{aegerter98}. The
same FLL order parameter vanishes continuously in that case, consistent
with a second order phase transition.

We have furthermore studied the effects of anisotropy by applying the field
at an angle to the $c$-direction. The melting temperatures at constant
field are consistent with the expectations of the scaling theory of
Blatter {\em et al.} \cite{blatterani}. A determination of the anisotropy
$\gamma$ however is not possible, as the deviations of a cos ($\theta$)
dependence are negligible in the range of angles studied. The nature of
the transition is unchanged by having the field at an angle, consistent
with the findings of Schilling {\em et al.} \cite{schilling97}, who
studied the dependence of the entropy change on angle.

The temperature dependence of the melting line is consistent with that
expected from a Lindemann criterion \cite{houghton}, where the
temperature dependence of the penetration depth follows that of the 3D XY
model. The observed melting line extends down to very low fields (0.2 T),
where $\mu$SR still indicates a very sharp transition. The range of
applicability of this temperature dependence is
limited, as we have shown from the temperature dependence of the
scattered intensity at low fields, where thermal fluctuations do not play
an important role. For temperatures below 75 K the approximation becomes
invalid. We would like to note that this is a rather unexpectedly big
range of applicability of the critical exponent of the 3D XY model. Using
the independently measured penetration depth \cite{stu} and anisotropy
\cite{cam} for our sample, we have also determined the Lindemann number
from the melting line, which turns out to be c$_L$ = 0.173(7). This is in
very good agreement with the Lindemann number previously obtained from a
detailed $\mu$SR study of the melting line in BSCCO \cite{lee97}.

\section{Acknowledgements}
We would like to thank Michel
Bonnaud, Andreas Polsak, Pierre George (ILL) and Alex Amato (PSI) for
technical support. Financial support
from the EPSRC of the UK and the Swiss National Science Foundation is
gratefully
acknowledged. This work was funded in part by Oak Ridge National Laboratory,
which is managed by Lockheed Martin Energy Research Corporation under contract
No. DE-AC05-96OR22464 for the US DOE.

\bibliographystyle{prsty}

\end{multicols}
\end{document}